\documentstyle[epsfig,sprocl]{article}
\input{epsf.tex}
\font \fwork = cmssqi8

\begin{document}
\hspace*{-18pt}{\epsfxsize=80pt  \epsfbox{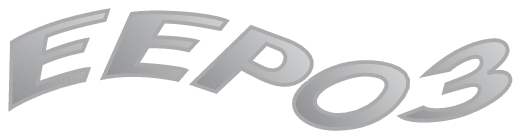}} 
\vspace*{-16pt}
\rightline{\fwork LPSC Grenoble, France, October 14-17, 2003}
\vspace*{15pt}
%
%
\title{\Large\bf Probing nucleon-nucleon correlations \\ with
  $\mathbf{(e,e'p)}$ and $\mathbf{(e,e'pp)}$ reactions}   

\author{J. Ryckebusch}

\address{
    {\it Department of Subatomic and Radiation Physics,
Ghent University, Belgium \\ http://inwpent5.UGent.be/}
        }
\maketitle

\begin{abstract}
Some of the recent attempts to detect signals of nucleon-nucleon
correlations with semi-exclusive $A(e,e'p)$ and exclusive $A(e,e'pN)$
processes are reviewed.  Unfactorized and distorted-wave calculations
for these processes are discussed.  The calculations implement
two-body currents stemming from pion-exchange and intermediate
$\Delta$ excitation.  The two-body currents are shown to be highly
competitive with the mechanisms related to nucleon-nucleon
correlations.  This observation seriously hampers attempts to link
semi-exclusive $A(e,e'p)$ measurements, which probe high missing
energies, to the correlated part of the nuclear spectral
function. Signals of central short-range correlations (SRC) have been
detected in $(e,e'pp)$ measurements on $^{12}$C and $^{16}$O, thereby
confirming the picture that in hadronic matter SRC solely affect
nucleon pairs with a small center-of-mass momentum and residing in a
relative $S$ state.
\end{abstract}

%
%
\section{Introduction}
\noindent

Two types of reactions involving electrons have been advocated as
potentially powerful tools to learn about that part of the nuclear
dynamics which cannot be understood as a mean-field phenomenon.
First, semi-exclusive $A(e,e'p)$ reactions, which probe the continuous
part of the spectral function, allow to study high-momentum protons in
kinematic conditions which favor the occurrence of highly correlated
nucleon pairs.  In interpreting the data, however, great care must be
exercised in evaluating the effect of competing processes, like
final-state re-scatterings, pion-production through intermediate
$\Delta$ creation and two-body meson-exchange currents.  Second,
triple-coincidence $A(e,e'pp)$ and $A(e,e'pn)$ measurements are
challenging and gained momentum from 1990 on-wards. Measurements of
two-proton knockout cross sections on the target nuclei $^3$He
\cite{groep2002,JLAB3He}, $^{12}$C \cite{raoul} and $^{16}$O
\cite{starink,guenther2000} performed at Mainz, Amsterdam and
Jefferson Lab, indicated important contributions from intermediate
$\Delta$ creation. Information about the short-range correlations
(SRC) can be extracted by comparing model calculations to the data.
These comparisons provided evidence that only diprotons residing in a
relative $S$ state and having a small center-of-mass (C.M.) momentum,
are subject to SRC.  The wave function describing the C.M. motion of
diprotons, on the other hand, was found to be completely in line with
mean-field predictions up to momenta of 0.5 GeV
\cite{raoul,watts2000}.  To date, attempts to measure $A(e,e'pn)$
cross sections aim at providing information about tensor correlations
in nuclei.

This contribution is organized as follows.  In Sect.~\ref{sec:general}
the use of correlation functions to implement nucleon-nucleon
correlations in reaction-model calculations will be sketched.
Sect.~\ref{sec:eepp} outlines a method of extracting information about
the correlation functions from measured two-nucleon knockout data.
Sect.~\ref{sec:eep} introduces an unfactorized model for computing
$A(e,e'p)$ cross sections at missing energies above the two-nucleon
emission threshold.

\begin{figure}[htb]
\centerline{\epsfig
{file=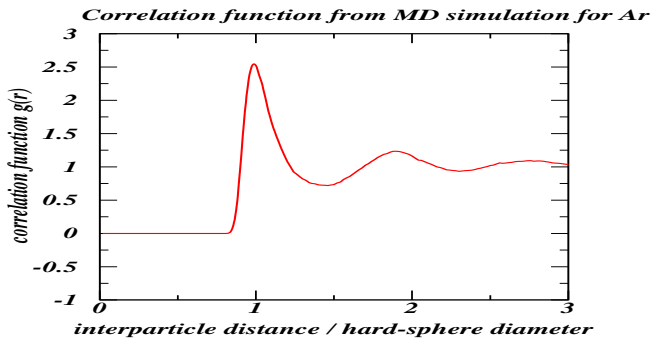,width=10.0cm,height=5.0cm}}
\caption{Typical correlation function from a molecular-dynamics (MD) simulation
of hard spheres interacting through a Van Der Waals
interaction. \label{fig:md} }
\end{figure}

\section{Correlations and electro-induced hadron knockout}
\label{sec:general}

An efficient way of modeling the ``fluctuations'' (or, beyond
mean-field behavior) of interacting many-body systems are the
correlation functions which in their simplest form are defined as
\begin{equation}
G({r}_1,{r}_2) = \left< {r}_1 \; {r}_2 \right>  - 
\left< {r}_1 \right> \left< {r}_2 \right>  \; .
\end{equation}
In most interacting quantum systems, the correlation function depends
solely on the relative coordinate ${r}_{12} = \mid \vec{r}_1 -
\vec{r}_2 \mid $.  In nuclear systems the correlation functions are
effectively operators due to the state dependence of the
nucleon-nucleon force.  The major fraction of the correlations in
hadronic matter can be incorporated economically by means of a
correlated wave function of the type \cite{walet}
\begin{equation}
\left| \widetilde{\Psi} \right> \equiv \frac {\widehat{\mathcal{G}} 
\left| \Phi \right> } 
{ \left< \Phi \left| \widehat{\mathcal{G}}^{\dagger}  \widehat{\mathcal{G}} 
\right| \Phi \right> }
\; ,
\end{equation} 
where $\Phi$ is a Slater determinant.  The correlation operator has a   
spin, isospin and tensor dependence
\begin{displaymath}
\widehat{\mathcal{G}}  = \widehat{\cal S} \left[ \prod _{i<j=1} ^{A} \left(
1 - {g_c(r_{ij})+f_{t\tau}(r_{ij})\widehat{S_{ij}} \vec{\tau}_i
. \vec{\tau}_j 
+ f_{\sigma \tau}(r_{ij}) \vec{\sigma}_i \vec{\sigma}_j  \vec{\tau}_i
. \vec{\tau}_j }
\right) \right]  \; ,\nonumber
\end{displaymath}
where $\widehat{\cal S}$ is the symmetrizing operator.  The
short-range correlations (SRC) are usually associated with the
so-called ``central'' correlation function $ g_c (\mid \vec{r}_1-
\vec{r}_2 \mid) $ and can be linked with the finite extension of the
nucleons which gives rise to a strong repulsion at short inter-nucleon
distances. In many respects, the central correlation function for
nucleons in nuclei can be expected to resemble those of molecules in
liquids.  As illustrated in Fig.~\ref{fig:md}, the latter exhibit a
fluctuating behavior as a function of the intermolecular distance.
Similarly, when moving with a nucleon in the nucleus, its finite
extension will induce a reduced probability of finding another nucleon
over distances of the order of its radius $R_p$ and an enhanced one at
distances slightly larger than $R_p$.  The limited spatial extension
of a nucleus makes the fluctuating behavior of $g_c(r_{12})$ to fade
out at inter-nucleon distances $r_{12}$ of a few femtometers.

Experimentally determining the correlation functions for nuclei turns
out to be challenging.  SRC effects are notoriously difficult to chase
with the electromagnetic probe \cite{riska96}.  Indeed, the continuity
equation
\begin{equation}
\vec{\nabla} _{\vec{r}} \cdot \vec{J} \left( \vec{r} \right) +
\frac {1} {i \hbar} \biggl[ \rho \left( \vec{r} \right), \hat{H}
\biggr] =0 \; \; \textrm{with,} \; \; \;
\hat{V}(1,2) = f( r_{12} ) \vec{\sigma} _1 \cdot \vec{\sigma} _2
\; \; \vec{\tau} _1 \cdot \vec{\tau} _2  \; ,
\end{equation}
demands that the corresponding $\vec{J}^{[2]}$ obeys
\begin{equation}
\vec{\nabla}  \cdot  \vec{J} ^{[2]}  
\left( \vec{r}_1, \vec{r}_2 ; \vec{r} \right)
 =  - \mid e \mid \; \; f( r_{12} ) 
\vec{\sigma} _1 \cdot \vec{\sigma} _2 
 \times  \Biggl[
\delta \left( \vec{r}  - \vec{r} _1 \right)  
- \delta \left( \vec{r}  - \vec{r} _2 \right) \Biggr] \left( \vec{\tau}_1
\times \vec{\tau _2} \right) _z  \; .
\end{equation}
At short ranges one has that $ \vec{r} _1 \longrightarrow \vec{r} _2 $
and $\vec{\nabla} \cdot \vec{J} ^{[2]} \approx 0$.  Accordingly,
exchange currents are dominated by the ``longer-range'' pion exchanges
and ``short-range'' phenomena tend to hide themselves for the electromagnetic
probe.

\section{SRC and $A(e,e'NN)$ reactions}
\label{sec:eepp}

The link between SRC and the two-nucleon knockout $A(e,e'NN)$ cross
sections can be easiest seen in the so-called factorized approach.  In
the spectator approximation this limit can be reached by describing
the emerging nucleons in terms of plane-waves and assuming that the
reaction occurs at short relative distances $\mid \vec{r}_{12} \mid
\approx 0$.  Then, the cross section takes on the formal form
\begin{equation}
\frac {d^8 \sigma} {d \epsilon ' d \Omega _{\epsilon '} d \Omega _{1} 
 d \Omega _{2} dT _{p_{2}}} (e,e'N_1 N_2)
       = E_1 p_1 E_2 p_2 f_{rec} ^{-1} 
{\sigma _{eN_{1}N_{2}} \left( 
              k _{+} , k  _{-} ,q        \right) }
{F_{h_{1},h_{2}}(P)} \; ,
\end{equation} 
where the relative momentum $\vec{p}_{rel}$ and C.M. momentum
$\vec{P}$ of the pair is 
\begin{equation}
 \vec{p}_{rel}  =  \vec{k}_{\pm} = \frac {\vec{p}_{1} -\vec{p}_{2}} {2}  
\pm \frac {\vec{q}} {2} \; \; \; \;  \; \; \; \;  \; \; \; \;  \; \; \; \;  
\vec{P}=\vec{p}_{1} + \vec{p}_{2} - \vec{q}  \; .
\end{equation}
In these expressions, $\vec{p}_1$ and $\vec{p}_2$ are the measured
momenta of the ejectiles.  In a naive spectator approach, the quantity
$P$ corresponds with the C.M. momentum of the diproton at the moment
that it is hit by the virtual photon.  In an ideal world with
vanishing FSI mechanisms, $p_{rel}$ denotes the two possible values
for the relative momentum of the bound pair.  Further, $F_{h_{1},h_{2}}(P)$
is the combined probability to find a dinucleon with C.M. momentum $P$
in the quantum state defined by two single-particle levels $(h_1,h_2)$
and $\sigma _{eN_{1}N_{2}}\left(k_{+},k_{-},q\right)$ is the
probability to have an electromagnetic interaction with a dinucleon
with relative momentum $k_{\pm}$. For the $A(e,e'pp)$ case, an
analytical expression for $\sigma
_{eN_{1}N_{2}}\left(k_{+},k_{-},q\right)$ has been derived
\cite{janplb}.  It accounts for two-body effects induced by $\Delta$
currents, central, tensor and spin-isospin correlations.  The
function $\sigma _{eN_{1}N_{2}}$ plays an analogous role as the
off-shell electron-proton $\sigma _{ep} ^{(CCx)}$ functions in
$A(e,e'p)$.

\begin{figure}[htb]
\centerline{\epsfig{file=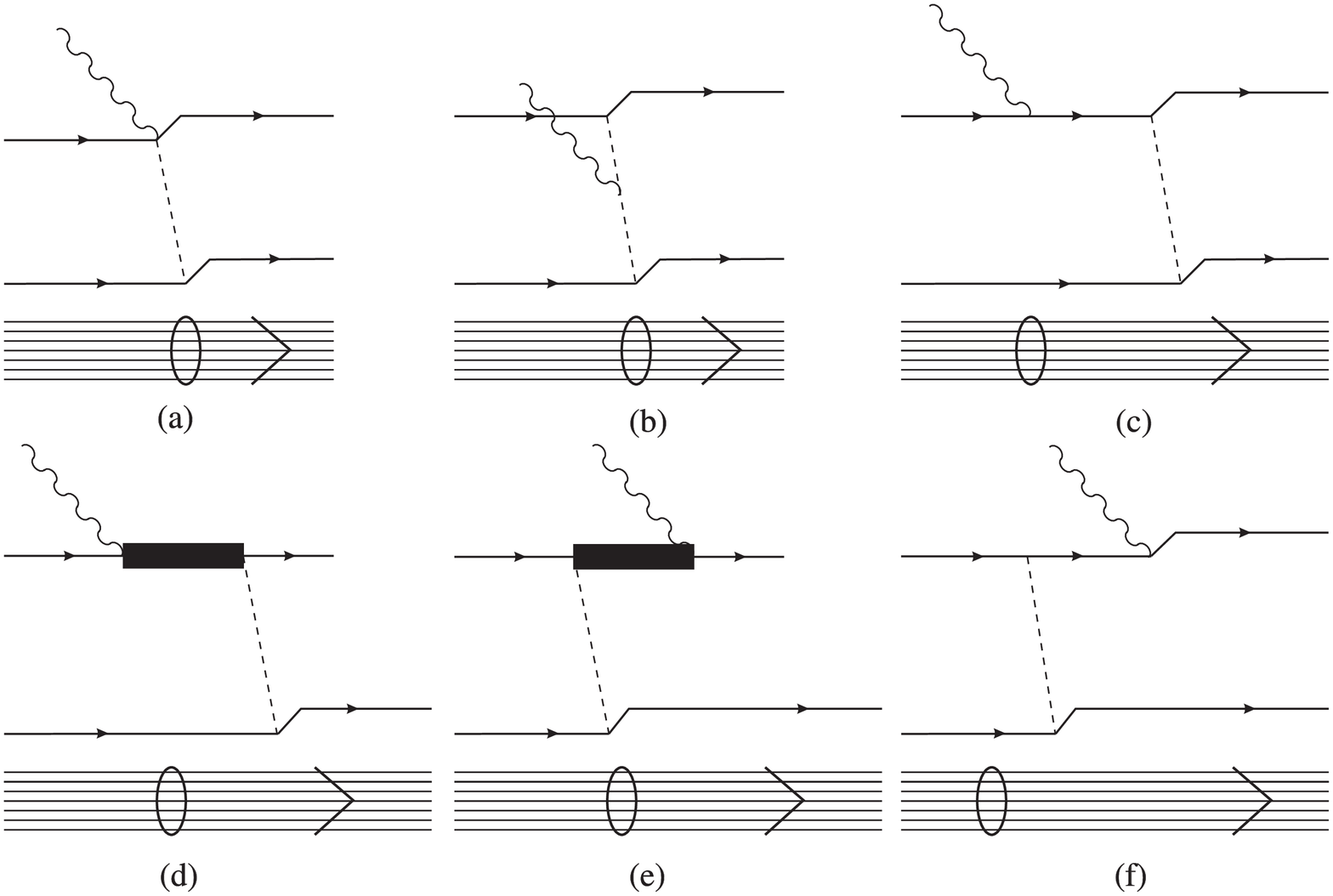,width=10.0cm,height=5.0cm}}
\caption{Typical diagrams included in a spectator model for
  electromagnetically induced two-nucleon knockout
  \label{fig:diagrams}.  Diagrams (a) and (b) are meson-exchange
  contributions.  Further, (c) and (f) represent the initial- and
  final-state correlations.  The $s$- and $u$-channel contribution for
  intermediate $\Delta$ creation and subsequent decay are (d) and (e).}
\label{fig:diagram}
\end{figure}

Various unfactorized microscopic models for computing A($e,e'NN$) (A
$\ge$ 12) observables have been developed over the last number of years
\cite{pavia,granada,gent}.  They all adopt the spectator approximation
and a distorted-wave description for the emerging
nucleons. Furthermore, differential cross sections can be computed for
each of the individual states in the final nucleus.  All models
include intermediate $\Delta$ excitation and SRC effects, but differ
in the way these mechanisms are implemented. A summary of the diagrams
which are usually implemented is displayed in Fig.~\ref{fig:diagram}.

The first high-resolution $A(e,e'pp)$ data which could clearly
separate the individual states in the final nucleus became recently
available \cite{guenther2000,kahrau1999}.  The data were collected by
the A1 collaboration with a unique three-spectrometer setup at the
MAMI facility in Mainz. An initial electron beam energy of 855~MeV and
an $^{16}$O target was used.  The two ejected protons, with momenta
$\vec{p}_1$ and $\vec{p}_2$, were detected parallel and anti-parallel
to the momentum transfer, a peculiar situation which is known as
``super-parallel kinematics''.  The energy and momentum transfer was
$<\omega>$=215~MeV and $<q>$=316~MeV.  Data were collected in a pair
C.M.  momentum range of $-100 \leq P \leq 400$~MeV/c.

\begin{figure}[htb]
\hspace{-1.cm}
{\epsfig{file=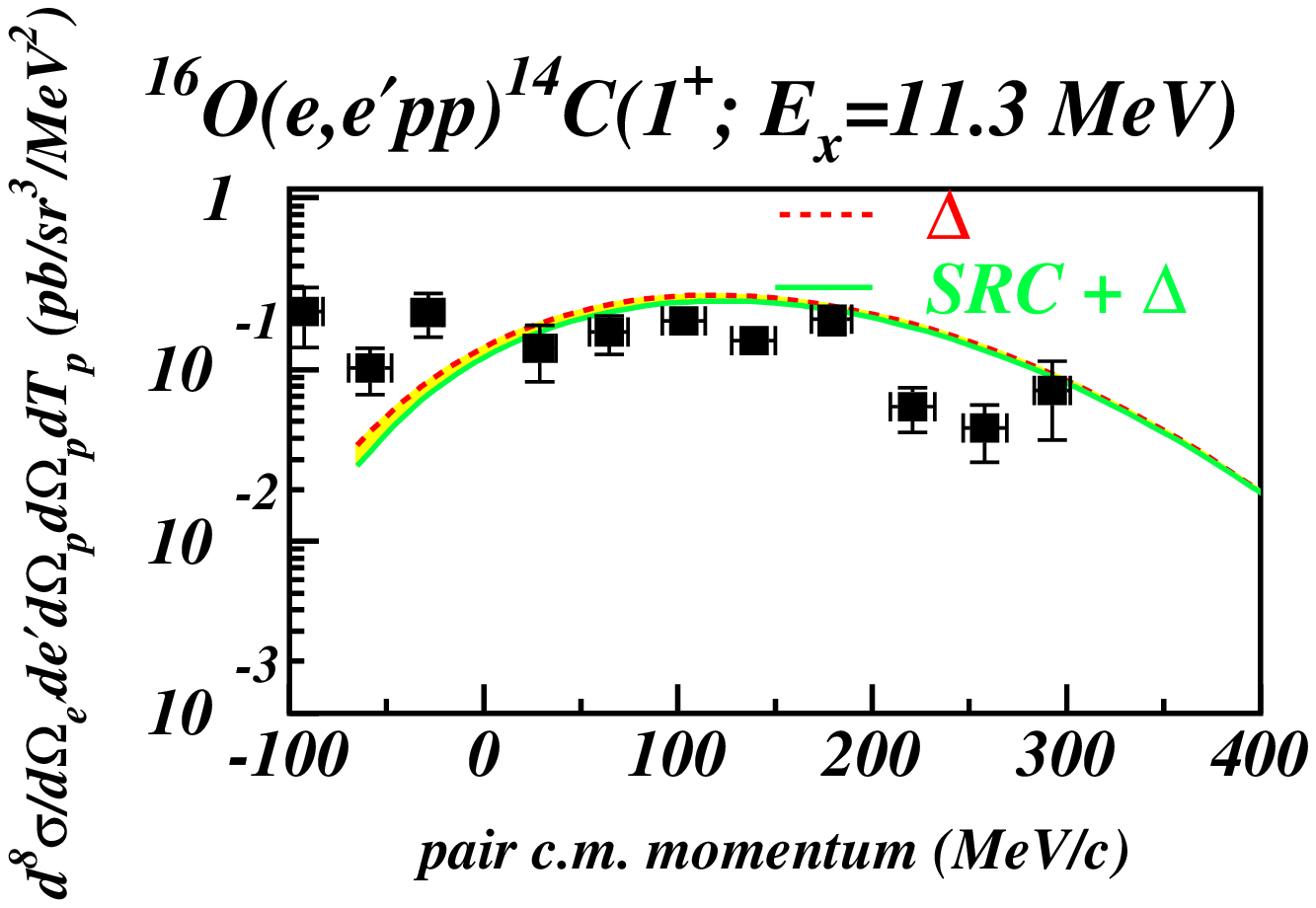,width=6.8cm}}
\hspace{-1.cm}
{\epsfig{file=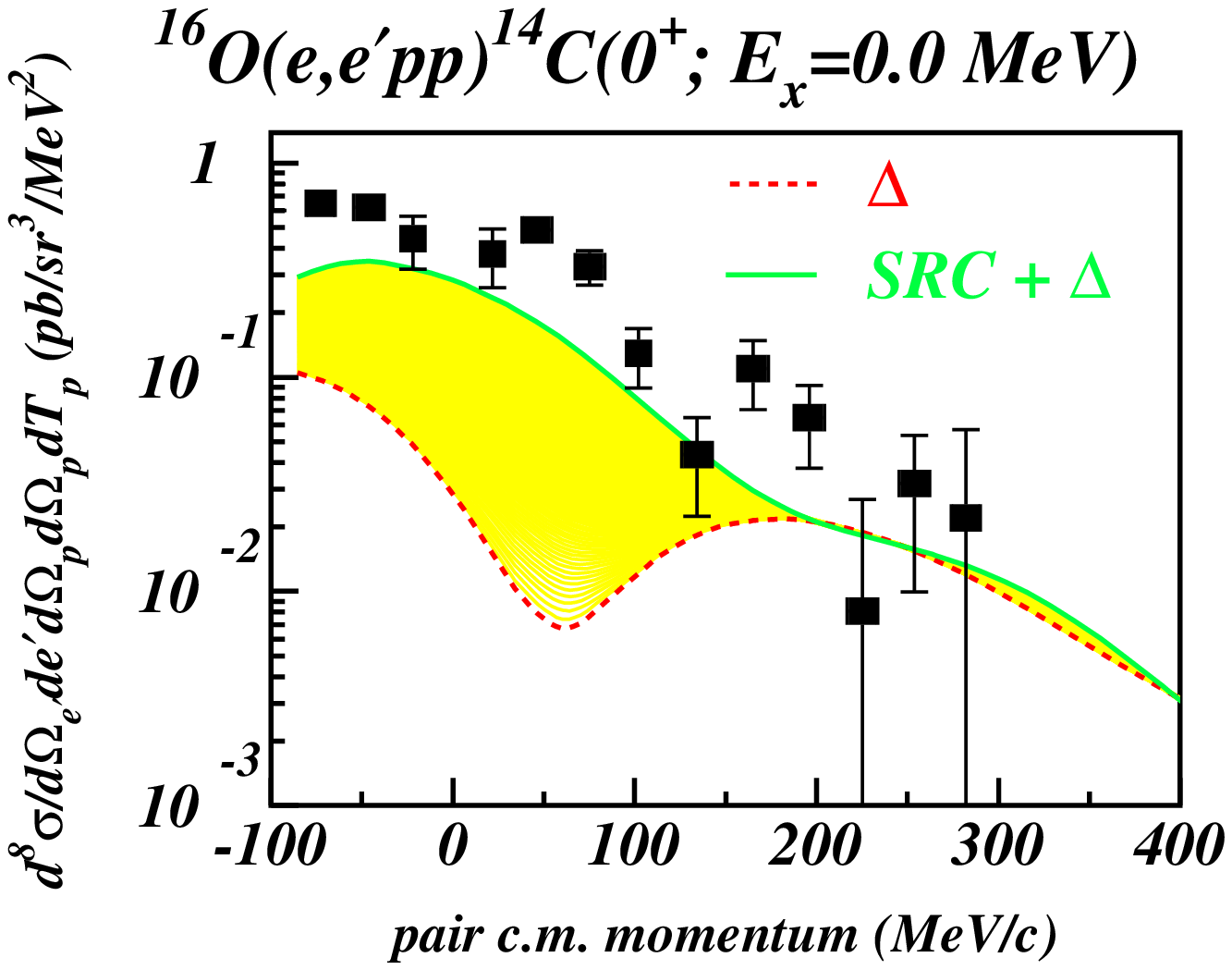,width=6.8cm}}
\caption{The eightfold differential cross section for the
$^{16}$O$(e,e'pp)^{14}$C$(0^+,E_x=0~MeV)$ and the
$^{16}$O$(e,e'pp)^{14}$C$(1^+,E_x=11.31~MeV)$ reaction as a function
of the pair C.M. momentum. The (red) dashed curve shows the results of the
distorted-wave calculations implementing only intermediate $\Delta$
excitation.  The (green) solid curve is the result of a
distorted-wave calculation that accounts for both
intermediate $\Delta$ and central short-range correlations. The data
are from Refs.~\protect \cite{guenther2000} and \protect
\cite{kahrau1999}. }
\label{fig:commainz}
\end{figure}

Figure~\ref{fig:commainz} presents a comparison between the Mainz
$^{16}$O$(e,e'pp)$ data of Ref.~\cite{guenther2000} and our
distorted-wave calculations for two specific states in the residual
nucleus.  More results can be found in Ref.~\cite{janepja}.  We start
our discussion with the $1^+$ state at an excitation energy of
$E_x=$11.31~MeV in the residual nucleus $^{14}$C.  In most
nuclear-structure calculations, the two-proton overlap amplitudes for
this particular state are dominated by the $\left| \left( 1p_{3/2}
\right) ^{-1} \left( 1p_{1/2} \right) ^{-1} ; 1^+ \right>$ two-hole
configuration.  The Moshinsky transformation serves as a guide to
identify the dominant relative and C.M. quantum numbers of the
diprotons for a specific transition.  Indeed, the quantum numbers of
the final state impose strong restrictions on the possible
combinations for the relative and C.M. angular momentum of the active
diproton. For the $\left| (1p)^{-2} ; 1^+ \right>$ configuration only
the combination of $L=1$ C.M. and $P$-wave relative wave functions for
the diproton is allowed.  A striking feature of the calculations for
the $1^+$ state displayed in Fig.~\ref{fig:commainz} is that the
intermediate $\Delta$ diagrams prevail in the computed angular cross
sections, while SRC effects are marginal.  Figure \ref{fig:commainz}
displays also a comparison of the
$^{16}$O$(e,e'pp)^{14}$C$(0^+,E_x=0~MeV)$ data and our reaction model
calculations.  For the ground-state transition, the distorted-wave
calculations including short-range correlations reproduce the
C.M.-momentum dependence well, while underestimating the data by
roughly a factor of two over the whole momentum range.  An interesting
observation from Fig.~\ref{fig:commainz} is that the calculation
ignoring SRC effects, underestimates the data for the ground-state
transition at low pair C.M. momenta by several factors.  At high pair
C.M. momenta the $L=1$ C.M. wave dominates and the calculations
neglecting SRC move closer to the data.  In any case, without
inclusion of central short-range correlations, neither the shape nor
the magnitude of the data for the ground-state transition can be
reproduced. We interpret this as strong evidence for short-range
correlations for proton pairs residing in relative $^1S_0$ states.  At
the same time, and equally important, central short-range correlations
appear to affect exclusively proton pairs residing in a relative $S$
state and having a small C.M. momentum.  This corresponds with the
picture that short-range correlations exclusively affect nucleon pairs
when they reside close to each other, thereby making them to move with
equal momentum in a back-to-back situation.

\begin{figure}
\begin{center}
{\mbox{\epsfysize=\textwidth\epsffile{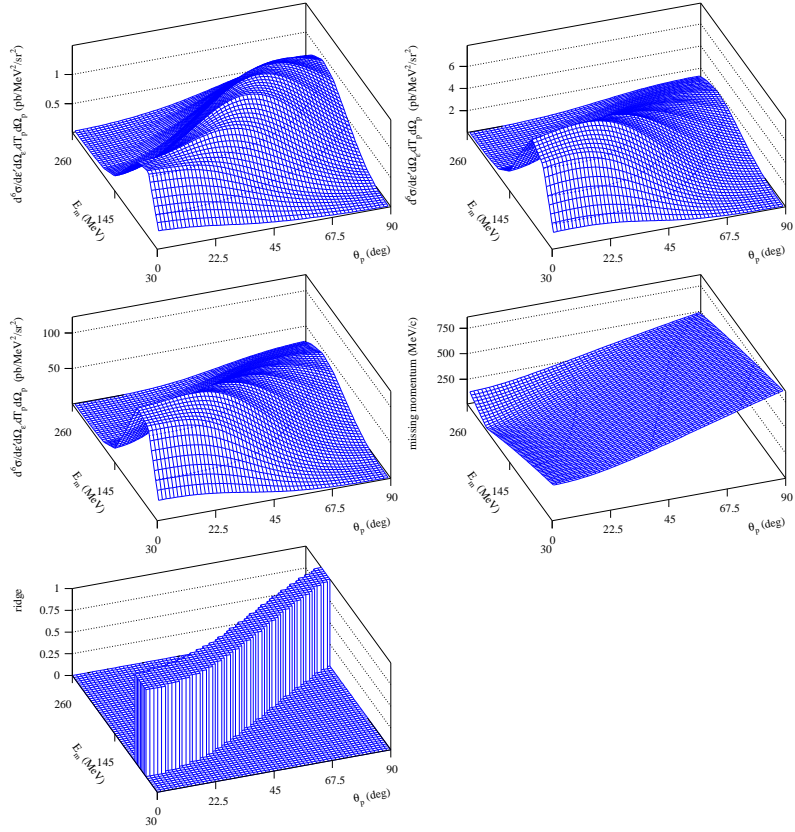}}}
\end{center}
\caption{The differential $^{16}$O$(e,e'p)$ cross section versus
missing energy ($E_m$) and proton angle ($\theta _p$) at
$\epsilon$=1.2~GeV, $\epsilon '$=0.9 ~GeV and $\theta_e$=16$^o$ (or,
$x \approx 0.15 $ and $q = 0.42$~GeV/c). The upper left panel includes
solely the central correlations and the upper right panel has both
central and tensor correlations.  The middle left panel includes apart
from the central and tensor correlations also the MEC and IC.  The
variation of $p_m$ versus missing energy and proton angle is shown in
the middle right panel. The lower left panel shows the position of the
``ridge'' (Eq. \protect \ref{eq:ridge}) in the $(E_m,\theta _p)$
plane.  Hereby, the variable $<E_x^{hh'}>$ was varied between 0. and
40.~MeV.}
\label{fig:central}
\end{figure}

%
\section{SRC and semi-exclusive $A(e,e'p)$ reactions}
\label{sec:eep}

Often, the semi-exclusive $A(e,e'p)$ measurements \cite{rohe2003}
which probe high missing energies ($E_m$) and momenta ($p_m$) are
interpreted starting from the following \textsc{factorized} expression
\begin{equation}
\frac {d^6 \sigma} {d T_p d \Omega _p d \epsilon ' d \Omega _{\epsilon
'}} (e,e'p)  =
\frac {p_p E_p} {(2\pi)^3}  \sigma _{ep} ^{CC1} 
P \left( \vec{p}_m, E_m \right) \; ,
\label{eq:eepfactor}
\end{equation}        
which assumes a direct connection between the correlated part of the
spectral function $P \left( \vec{p}, E \right)$ and the measured
differential cross sections.  The above expression which connects the
measured differential cross section to the spectral function relies on
the assumption that single-nucleon currents prevail and that the
strength attributed to contaminating mechanisms like two-nucleon currents
and re-scattering effects is highly suppressed.  The two-nucleon
knockout investigations reported in previous section, however, nicely
illustrated the importance of two-nucleon meson-exchange (MEC) and
isobar currents (IC) when chasing nucleon-nucleon correlations in
nuclei. In Ref.~\cite{stijncor} an \textsc{unfactorized} framework for
computing semi-exclusive $A(e,e'p)$ cross sections is proposed.  It
postulates that single-nucleon knockout to the energy continuum in the
(A-1) nucleus is a two-nucleon emission process whereby one of the
emerging nucleons remains undetected.  Relying on the model which was
developed for two-nucleon knockout, the angular cross sections are
determined through the following expression
\begin{equation}
\frac {d^6 \sigma} {d T_p d \Omega _p d \epsilon ' d \Omega_ {
\epsilon' } } {(e,e'p)} 
= {\sum _{N \equiv p,n} \int
d \Omega_N d E_{A-2}} \frac {d^9 \sigma} {d T_N d \Omega _N d T_p d
\Omega _p d \epsilon ' d \Omega _{\epsilon '} } 
{(e,e'pN)} \; ,
\end{equation} 
which involves an integration over the phase space of the undetected
nucleon.  Our approach allows to compute the strength from the MEC and
IC in the same framework in which also the contributions from
ground-state correlations are determined.  At the same time, it
provides a unified description to $A(e,e'p)$ in the continuous part of
the spectrum and $A(e,e'NN)$, the two most popular reactions used to
detect signatures of nucleon-nucleon correlations.

\begin{figure}
\centerline{\epsfig{file=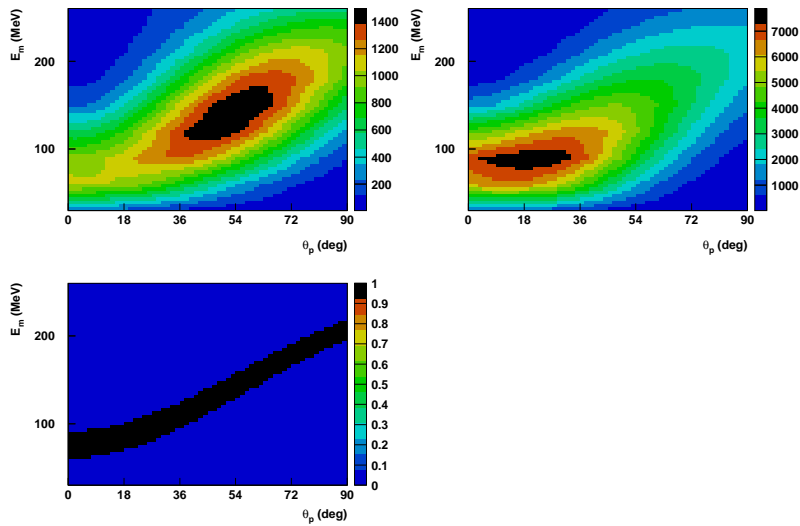,width=0.9\textwidth}}
\vspace{-4.5cm}
\caption{Contour plots corresponding with the upper and lower panels
  of Fig.~4.}
\label{fig:contour}
\end{figure}

The $^{16}$O$(e,e'p)$ differential cross sections presented in
Fig.~\ref{fig:central} correspond with a small Bjorken scaling
variable $x \approx 0.15 $ and are obtained by incoherently adding the
separately computed $^{16}$O$(e,e'pn)$ and $^{16}$O$(e,e'pp)$
strengths.  Thereby, two-nucleon knockout from all $(1s_{1/2},
1p_{3/2}, 1p_{1/2})$ shell-combinations is included.  We first address
the issue how the central, tensor and spin-isospin correlations
manifest themselves in the $(e,e'p)$ differential cross sections at
high $E_m$. We display the differential $^{16}$O$(e,e'p)$ cross
section versus missing energy and polar angle (measured with respect
to the direction of the momentum transfer). The variation with $\theta
_p$ allows to study the $p_m$ dependence of the cross section at a
fixed $E_m$.  For the sake of convenience, a panel with the variation
of $p_m$ versus $E_m$ and $\theta _p$ was added.  Roughly speaking,
the probed missing momentum $p_m$ increases with increasing
$\theta_p$.  Comparing the two upper panels in Fig.~\ref{fig:central}
it emerges that the strength attributed to the tensor correlations
largely overshoots the strength from the central correlations.  In
particular, this holds for the small proton angles $\theta _p$.  At
these angles, typically the smallest missing momenta are probed.  The
effect of the spin-isospin correlations is at the few percent level.
The central correlations are observed to manifest themselves in a
wider range of the ($E_m,\theta_p$) plane than the tensor correlations
do.  Apparently, the effect of the tensor correlations is confined to
a region of relatively low and moderate missing momenta, whereas the
contribution from the central correlations extends to higher proton
angles $\theta _p$ where typically higher missing momenta are probed.
This qualitative behavior of the calculated differential cross
sections reflects the fact that central correlations are the most
important correlations in the spectral function at really high missing
momenta, whereas the intermediate range is usually dominated by the
tensor correlations.

\begin{figure}
\begin{center}
\vspace{-2.cm}
{\mbox{\epsfxsize=\textwidth\epsffile{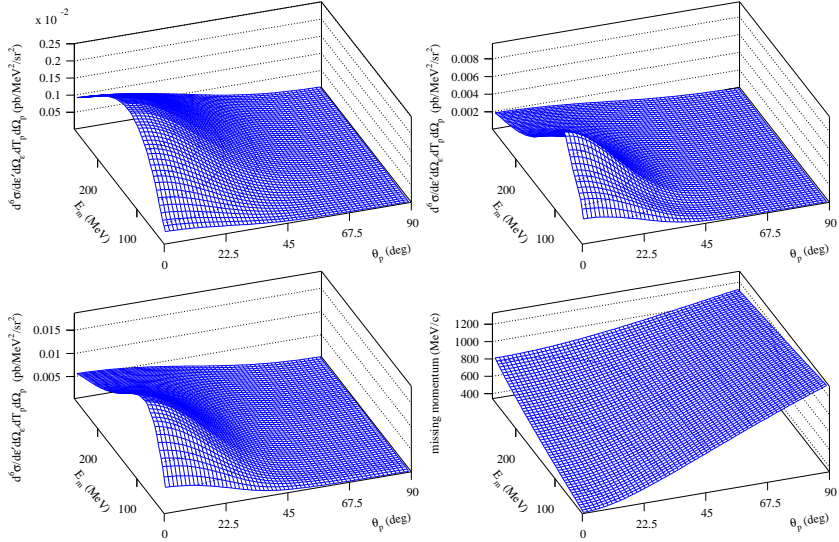}}}
\vspace{-1.5cm}
\end{center}
\caption{As in Fig. \protect \ref{fig:central} but now for kinematics
  corresponding with Bjorken x
  $\approx 2$, $\epsilon$=2.5 GeV and $Q^2$=1.1 GeV$^2$.} 
\label{fig:xis2}
\end{figure}

Another interesting feature of how the ground-state correlations
manifest themselves in the $(E_m,\theta _p)$ plane is that the peak of
the $^{16}$O$(e,e'p)$ differential cross sections shifts to higher
missing energies $E_m$ as one gradually moves to non-parallel
kinematics and higher $p_m$ values are probed.  This observation is a
manifestation of a well-known feature of the correlated part of the
spectral function, namely that the average missing energy $\left< E_m
\right>$ is predicted to increase quadratically in the missing
momentum according to
\begin{equation}
 \left< E_m
\right> = \frac {A-2} {A-1} \frac {p_m^2}{2M_N} + S_{2N} +
\left< E^{hh'}_{A-2} \right> \; ,
\label{eq:ridge}
\end{equation}
where $S_{2N}$ is the threshold energy for two-nucleon knockout and
$\left< E_{A-2}^{hh'} \right>$ the average excitation energy of the $A-2$ system
that was created after two nucleons escaped from the orbits
characterized by the quantum numbers $h$ and $h'$.  Moreover, the
strength in the correlated part of the spectral function is often
predicted to reside in a rather narrow region on both sides of $
\left< E_m \right>$ (the so-called ``ridge'' in the spectral function).  For the sake
of convenience, a panel with the exact location of the ridge in the
$(E_m,\theta_p)$ plane was added to Fig.~\ref{fig:central} and the
contour plots corresponding with the upper and lower panels of
Fig.~\ref{fig:central} are shown in Fig.~\ref{fig:contour}.  Indeed,
the major part of the calculated strength is concentrated in a wide
band of missing energies about this ridge.  Despite the fact that our
calculations are unfactorized, the above observations with
regard to the qualitative behavior of the calculated differential
cross sections, illustrate that they exhibit the
same qualitative features than what could be expected to happen in a
factorized approach based upon the Eq.~(\ref{eq:eepfactor}) using a
realistic ``correlated'' spectral function.

Also the contribution of MEC and IC to the $^{16}$O$(e,e'p)$ reaction
at high $E_m$ is shown in Fig.~\ref{fig:central}.  For the kinematics
considered there, they produce signals which are almost one order of
magnitude larger than the combined effect of central and tensor
correlations do.  It is important to note that also the two-body
currents create the major part of the $(e,e'p)$ strength along the
ridge (Eq.~(\ref{eq:ridge})) in the $(E_m, \theta _p)$ plane.
Therefore, the observation of a ``ridge'' should \textsc{not} always
lead one to conclude that signatures of nucleon-nucleon correlations
have been probed.  Indeed, contaminations from two-body currents will
feed similar parts of the $(e,e'p)$ phase space as the nucleon-nucleon
correlations do. The MEC and IC effects are confined to the transverse
part of the cross section, whereas the SRC mechanisms will feed both
the longitudinal and transverse part.  The observed transverse
enhancement of the $^{12}$C$(e,e'p)$ cross section at high $E_m$
\cite{dutta} can thus be naturally explained by the observations made
in Fig.~\ref{fig:central}.  The MEC and IC feeding of $A(e,e'p)$ at
high $E_m$ will gradually lose in importance with increasing $Q^2$
and/or kinematic conditions corresponding with Bjorken x $\ge$~1
\cite{stijncor}.  An example of what can be expected with more
favorable kinematic conditions corresponding with Bjorken $x \approx
2$ is displayed in Fig.~\ref{fig:xis2}.

%

\section{Conclusions}
Signatures of nucleon-nucleon correlations (or, beyond mean-field
behavior) are expected to reveal themselves in $A(e,e'p)$ at high
missing energy and momentum and $A(e,e'NN)$.  The $A(e,e'pp)$ data
acquired over the past decade has indeed revealed evidence for
short-range correlations, but indicated the importance of two-body
currents as a highly competitive reaction mechanism.  Tensor
correlations are predicted to produce far stronger signals than the
central short-range correlations and are presently under investigation
with the $A(e,e'pn)$ reaction.  Extracting clean information about the
nuclear dynamics originating from nucleon-nucleon correlations
requires the suppression of two-body current contributions.  In
$A(e,e'p)$ at high $E_m$ and $p_m$, such
conditions can be accomplished at $Q^2$ values exceeding 1~GeV$^2$ and
selective kinematics corresponding with Bjorken $x \approx 2$.

\vspace{0.5cm}

\noindent \textbf{References}

%


\begin{thebibliography}{99}
%
%
\bibitem{groep2002} D.L. Groep \textit{et al.}, Phys. Rev. C
\textbf{63} (2001) 014005.

\bibitem{JLAB3He} R.A. Niyazov \textit{et al.}, nucl-ex/0308013 and
  S. Gilad, contribution to this proceedings.

\bibitem{raoul}
K.I. Blomqvist \textit{et al.}, Phys. Lett.  \textbf{B421} (1998) 71.

\bibitem{starink} R. Starink {\em et al.}, Phys. Lett. {\bf B474}
(2000) 33.

\bibitem{guenther2000}
G. Rosner, Prog. Part. Nucl. Phys. {\bf 44} (2000) 99.

\bibitem{watts2000} D.P. Watts \textit{et al.}, Phys. Rev. C {\bf 62}
(2000) 014616.

\bibitem{walet} Niels R. Walet and R.F. Bishop, physics/0307069.

\bibitem{riska96}
D.O. Riska, Nucl. Phys. \textbf{A606} (1996) 251.

\bibitem{janplb}
Jan Ryckebusch, Phys. Lett. \textbf{B383} (1996) 1.

\bibitem{pavia}
C. Giusti, F.D. Pacati, K. Allaart, W.J.W. Geurts,
  W.H. Dickhoff and H. M\"{u}ther, Phys. Rev. C \textbf{57} (1998)
  1691.

\bibitem{granada}
Marta Anguiano, Giampaolo C\'{o} and Antonio M. Lallena,
J. Phys. G \textbf{29} (2003) 1119.


\bibitem{gent} J. Ryckebusch, M. Vanderhaeghen, L. Machenil and
M. Waroquier, Nucl. Phys. {\bf A568} (1994) 828.




\bibitem{kahrau1999} Marco Kahrau, Ph.D. thesis ``Untersuchung von
Nukleon-Nukleon Korrelationen mit Hilfe der Reaktion
$^{16}$O$(e,e'pp)^{14}$C in super-paralleler Kinematik'',
(Johannes Gutenberg-Universit\"{a}t, Mainz, 1999). URL : 
http://wwwa1.kph.uni-mainz.de/A1/publications/doctor/kahrau.ps.gz

\bibitem{janepja} J. Ryckebusch and W. Van Nespen, nucl-th/0312056.

\bibitem{rohe2003} D. Rohe, Eur. Phys. J. A {\bf 17} (2003) 439 and
  contribution to this proceedings.

\bibitem{stijncor} 
Stijn Janssen, Jan Ryckebusch, Wim Van Nespen and Dimitri Debruyne
Nucl. Phys. \textbf{A672} (2000) 285.

\bibitem{dutta}
D. Dutta \textit{et al.}, Phys. Rev. C \textbf{61} (2000) 061602.





%
\end{thebibliography}
\end{document}